# Giant polarization and abnormal flexural deformation in bent freestanding perovskite oxides

Songhua Cai[1,7,8], Yingzhuo Lun[2,8], Dianxiang Ji[1,8], Lu Han[1], Changqing Guo[2], Yipeng Zang[1], Si Gao[1], Yifan Wei[1], Min Gu[1], Chunchen Zhang[1], Zhengbin Gu[1], Xueyun Wang[2], Christopher Addiego[5], Daining Fang[3,4], Yuefeng Nie[1,*], Jiawang Hong[2,*], Peng Wang[1,*], Xiaoqing Pan[5,6,*]

[1] National Laboratory of Solid State Microstructures, Jiangsu Key Laboratory of Artificial Functional Materials, College of Engineering and Applied Sciences and Collaborative Innovation Center of Advanced Microstructures, Nanjing University, Nanjing 210093, China

[2] School of Aerospace Engineering, Beijing Institute of Technology, Beijing 100081, China

[3] Institute of Advanced Structure Technology, Beijing Institute of Technology, Beijing 100081, China

[4] State Key Laboratory for Turbulence and Complex Systems & Center for Applied Physics and Technology, College of Engineering, Peking University, Beijing 100871, China

[5] Department of Physics and Astronomy, University of California-Irvine, Irvine, California 92697, USA

[6] Department of Materials Science and Engineering, University of California-Irvine, Irvine, California 92697, USA

[7] Department of Applied Physics, The Hong Kong Polytechnic University, Hung Hom, Kowloon 999077, Hong Kong

[8] These authors contributed equally to this work.

*Correspondence to: E-mail: wangpeng@nju.edu.cn; xiaoqing.pan@uci.edu; hongjw@bit.edu.cn; ynie@nju.edu.cn

**Recent realizations of ultrathin freestanding perovskite oxides offer a unique platform to probe novel properties in two-dimensional oxides. Here, we observed a giant flexoelectric response in freestanding BiFeO$_3$ and SrTiO$_3$ in their bent state arising from strain gradients up to $4\times10^7$ m$^{-1}$, suggesting a promising approach for realizing extremely large polarizations. Additionally, a substantial reversible change in thickness was discovered in bent freestanding BiFeO$_3$, which implies an unusual bending-expansion/shrinkage and thickness-dependence Poisson's ratios in this ferroelectric membrane that has never been seen before in crystalline materials. Our theoretical modeling reveals that this unprecedented flexural deformation within the membrane is attributable to a flexoelectricity–piezoelectricity interplay. The finding unveils intriguing nanoscale electromechanical properties and provides guidance for their practical applications in flexible nanoelectromechanical systems.**

Electromechanical properties of functional materials play a significant role in electronic devices[1,2]. For perovskite oxides with unique electromechanical functionalities, strain engineering can stimulate significant electronic phenomena[3-5], such as inducing polarization in the nonpolar material SrTiO$_3$ and attaining a record-high polarization value in PbTiO$_3$[6,7]. In contrast to a homogenous strain, a strain gradient is inversely proportional to the spatial scale and can increase by seven orders of magnitude when the system shrinks from macroscale (~1/m) to nanoscale ($10^7$/m)[8]. Therefore, flexoelectricity, the coupling between polarization and strain gradient, becomes a significant and even dominant effect at this nanoscale, and hence, potentially induces novel physical phenomena that have attracted much attention[9-11]. However, the simultaneous implementation of strain and strain-gradient engineering is traditionally subject to the specifications of substrates or epitaxial conditions, which largely restrict the tunability and scalability of strains and strain gradients. The recently developed sacrificial buffer-layer technique using water-soluble $Sr_3Al_2O_6$ (SAO) provides a reliable method to synthesize high-quality freestanding thin perovskite oxides[12]. The structural stability of these perovskite oxides such as $BiFeO_3$ (BFO) and $SrTiO_3$ (STO) has been demonstrated in previous work[13]. Benefitting from the excellent flexibility of these oxides[14], nanoscale mechanical bending offers a new approach in strain and strain-gradient engineering. Given that the freestanding oxides are only a few nanometers thick, they are able to generate a huge strain gradient during bending, which may induce an ultrahigh polarization. Additionally, these unconventional low-dimensional systems may stimulate other novel physical and mechanical responses via electromechanical coupling effects (i.e., piezoelectricity and flexoelectricity)[8], similar to their bulk counterparts, which are known to exhibit a multitude of physical properties[15].

In addition to the novel strain-induced electrical properties of perovskite oxides[16,17], recent studies have found some interesting mechanical properties arising from the interplay between piezoelectricity and flexoelectricity[18-20] which implies that mechanical responses can be modulated via strain gradients. With this interplay in freestanding ultrathin films (several-unit-cell thickness), more interesting mechanical phenomena are expected that may even require a rethink of classical elasticity theory of materials in the presence of a huge strain gradient ($\sim 10^7$ m$^{-1}$) at small scales.

In this work, we report huge polarization enhancements in high-quality flexible freestanding perovskite oxides of polar BFO and nonpolar STO subject to ultra-high strain gradients up to $4\times 10^7$ m$^{-1}$, revealing that flexoelectricity plays a dominant role in determining the polarization at nanoscale. More interestingly, in addition to this giant polarization in BFO membranes, our results uncover unusual mechanical properties featuring bending-expansion/shrinkage and a hyperbolic-like distribution in Poisson's ratio. This is beyond the realm of classical elasticity theory, in which thickness and Poisson's ratio are assumed constant in bent single-phase membranes. Furthermore, a theorical model reveals these irregular mechanical phenomena to be governed by an interplay between flexoelectricity and piezoelectricity. Our results expose novel electrical and mechanical properties in bent freestanding perovskite oxides. Moreover, a new area of nanoscience is opening up allowing the tuning of electromechanical behaviors via giant strain gradients at the atomic scale that is crucial in the related research and potential applications of nano-electromechanical systems.

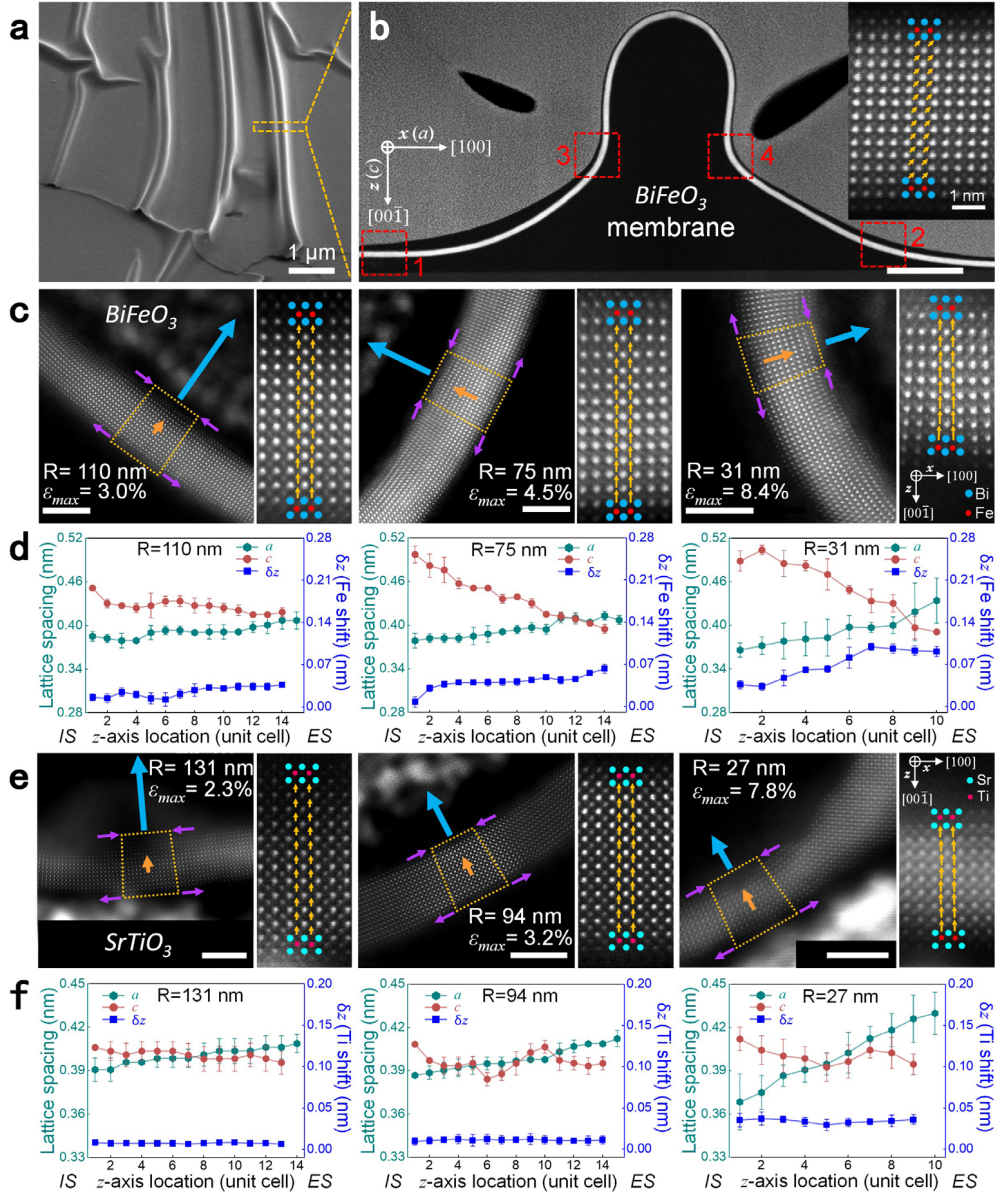

**FIG. 1. Bending behavior of BFO and STO membranes. a,** Scanning electron microscope (SEM) image of wrinkled freestanding BFO. Scale bar: 1 μm. **b,** Cross-sectional STEM-HAADF image of a single wrinkle in (**a**). Scale bar: 100 nm. Inset indicates the unstrained state taken from region 1 (marked by red square), where the spontaneous polarization has an upward out-of-plane component (yellow arrow). **c,** STEM-HAADF images taken from the bent regions 2 (left), 3 (middle) and 4 (right) in (**b**). R represents the radius of curvature. The lattice begins rupturing from the external surface under 8% strain. Blue and yellow arrows indicate bending directions and the polarization distribution, respectively. Purple arrows represent the in-plane strain. Scale bar: 5 nm. **d,** Line profiles of the lattice spacings $c$ and $a$, and the displacements of the Fe atomic columns $\delta z$ across from the internal (IS) to external surfaces (ES) in (**c**). **e,** STEM-HAADF images of freestanding STO under 2%, 3%, and 8% strain. Under 8% strain, the lattice structure of internal surface begins to rupture. Scale bar: 5 nm. **f,** Line profiles of lattice spacings $c$ and $a$, and the displacements of the Ti atomic columns $\delta z$ across from IS to ES in (**e**).

Freestanding ultrathin BFO and STO (~6 nm thickness) with corrugations were fabricated (Fig. 1a) and further prepared as cross-sectional transmission electron microscopy (TEM) samples. Atomic-resolution scanning TEM high-angle annular dark-field (STEM-HAADF) images of BFO (Fig. 1c) and STO (Fig. 1e) were acquired from different strained regions. The lattice structures have maintained their integrity and continuity without any obvious rupture even under a giant strain up to 8% in both BFO and STO, indicating the high flexibility of freestanding thin perovskite oxides. We observed a huge strain gradient arising from a considerable change in the in-plane strain corresponding to variations in lattice spacings $a$ (Fig. 1d, f) across the membrane in the nanometer range. Here, we call the surface of a bent membrane facing toward the center of curvature (Fig. 1c, e) the internal surface (IS), which is subject to compressive in-plane strain (negative values), and the opposing surface the external surface (ES), which is subject to tensile in-plane strain (positive values). The maximum strain gradients of the BFO and STO membranes are up to ~$3.7 \times 10^7$ m$^{-1}$ and $4.3 \times 10^7$ m$^{-1}$ respectively, which are nearly one order of magnitude larger than those generated in epitaxial films ($10^5$–$10^6$ m$^{-1}$)[9,10,21]. Although similar strain gradients were observed just near the tip of an atomic force microscope ($10^6$–$2 \times 10^7$ m$^{-1}$)[22,23], the uniformity of this huge strain gradient across the entire thickness range of the film is a unique property of the bent membranes which also offers better tunability in flexoelectric applications.

The polarization evolution (marked by yellow arrows in Fig. 1c, e) in BFO and STO membranes are associated with lattice distortions. To understand the relationship between this enhanced mechanical strain/strain gradient and the corresponding polarization at the nanoscale, quantitative probing of the polarization at single unit cell level is essential but remains challenging. Here, we used a widely accepted method assuming that the polarization in $AB$O$_3$ perovskites is proportional to the off-centering displacement of the B site atom ($\delta z$) with respect to the center of four surrounding A site atoms[24-26]. A recently developed differential phase contrast method was applied to calibrate the polarization measurement[27] and indicated that the former method is able to provide relative magnitudes and variational trends of polarizations faithfully for the following analysis.

In ferroelectric BFO without bending, we found the off-centering displacement of Fe ions ($\delta z$-Fe) to be oriented diagonally and the magnitude of the out-of-plane displacement to be nearly uniform across the membrane (inset in Fig. 1b). However, after bending, this displacement is oriented along the thickness direction and decreases from the ES (tensile in-plane strain) to the IS (compressive in-plane strain); see Fig. 1c, d. For a bent freestanding STO, the Ti ions also undergo a displacement ($\delta z$-Ti) arising from flexoelectricity and increases with decreasing radius of curvature during bending (Fig. 1e, f). Interestingly, unlike the BFO case, displacement $\delta z$-Ti remains almost constant cross the membrane, instead of decreasing from the ES to the IS.

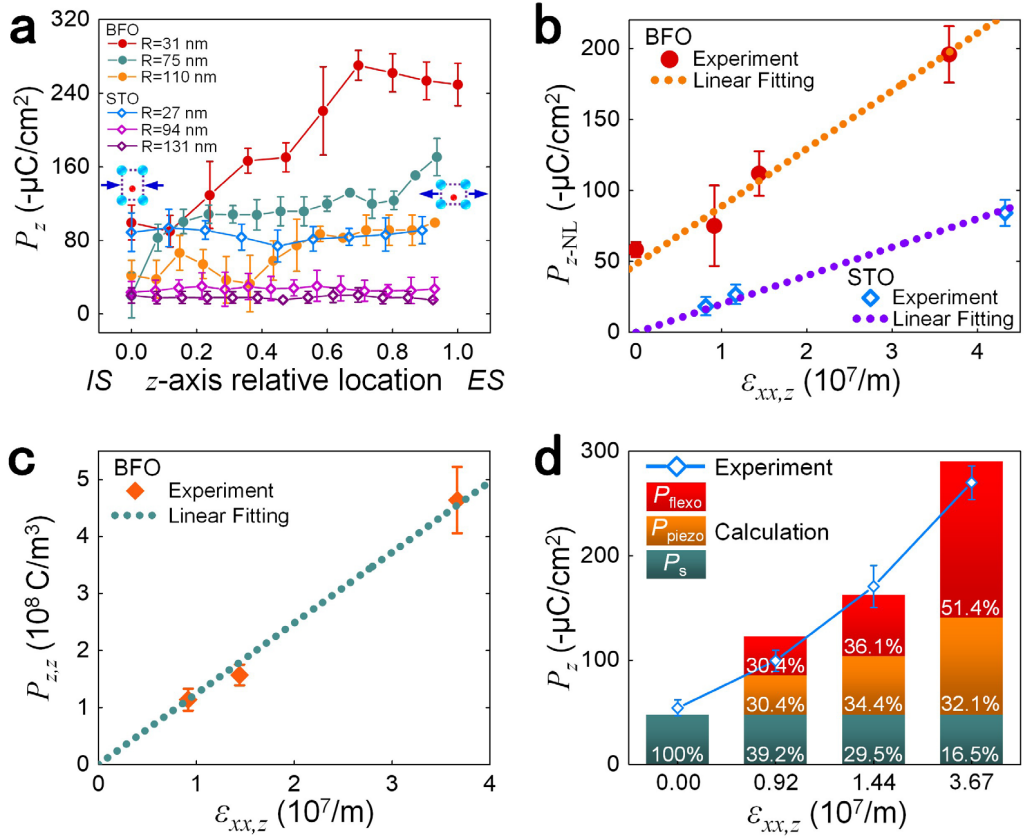

**FIG. 2. Flexoelectric enhancement of membranes polarization. a,** Out-of-plane polarization varied with $z$-axis relative locations from the internal (IS) to the external surfaces (ES). **b,** Linearity in the out-of-plane polarizations at neutral layers and their corresponding strain gradients. **c,** Linearity in the out-of-plane polarization gradient and corresponding strain gradients. **d,** Calculated proportions of different components (flexoelectricity, piezoelectricity and spontaneous polarization) contributing to the maximum polarization experimentally measured at outermost layers of bent BFO with three different strain gradients.

With the B-site displacement measured from the STEM-HAADF images, we obtained the polarization distribution in bent freestanding BFO and STO membranes (Fig. 2a). Several features were noted: 1) the smaller the radius of curvature from bending, the larger the strain and strain gradient that subsequently produce larger polarizations in both BFO and STO; 2) the polarization distribution is nearly constant for STO but in BFO increases from the IS to the ES (the $c/a$ ratio decreases from IS to ES because of a decrease (increase) in lattice spacing $c$ ($a$)); and 3) the maximum polarization occurs in a bent membrane with a minimum radius of curvature and possessing the largest strain and strain gradient. For example, for the 8.4%-strained freestanding BFO with the smallest radius of 31 nm (strain gradient up to $3.7\times10^7$ m$^{-1}$), the maximum polarization reached was approximately 250±23.2 μC/cm$^2$ in magnitude at the external layers (Fig. 2a). This magnitude is significantly larger than that in a tetragonal-like BFO film (130 μC/cm$^2$)[28]. Note also that the giant polarization obtained here is even comparable to that acquired in PbTiO$_3$ (236 μC/cm$^2$), which is the largest polarization found in ferroelectric materials so far[7]. For bent STO with a radius of curvature of 27 nm (strain gradient up to $4.3\times10^7$ m$^{-1}$), its polarization magnitude reaches ~86±6.1 μC/cm$^2$, which is also larger than that found recently in a STO crack tip and attributable to flexoelectricity (62±16 μC/cm$^2$)[29].

This giant polarization observed in these bent freestanding perovskite oxides is attributed to both piezoelectricity and flexoelectricity; in particular, the latter provides a major contribution in such extreme strain-gradient conditions. The total out-of-plane polarization $P_z$ origins from out-of-plane spontaneous $P_{s\perp}$, piezoelectric and flexoelectric polarizations; that is, $P_z = P_{s\perp} + \tilde{e}_{zxx}\varepsilon_{xx} + \tilde{\mu}_{zxxz}\varepsilon_{xx,z}$, where $\tilde{e}_{zxx}$ and $\tilde{\mu}_{zxxz}$ are the coefficients of the effective transverse piezoelectricity and flexoelectricity, and $\varepsilon_{xx}$ and $\varepsilon_{xx,z}$ are the in-plane strain and its gradient along the $z$ direction. In the calculations of polarizations and strain gradients throughout this work, we define the [00$\bar{1}$] crystal direction of the membranes as the positive direction (Fig. 1).

For simplification, the neutral layers where the strain is zero ($\varepsilon_{xx} = 0$) are chosen in the analysis of $P_z$ and the strain gradient. The neutral layer polarization $P_{z-NL}$ changes linearly with the strain gradient (Fig. 2b); the slopes of the fitted lines are the coefficients of flexoelectricity $\tilde{\mu}_{zxxz}$ and found to be $-40.7 \pm 6.3$ nC/m and $-20.0\pm0.9$ nC/m for BFO and STO, respectively (the negative sign indicates that the direction of the flexoelectric polarization is opposite to the strain gradient and always points towards the center of curvature). The magnitude of these coefficients matches well with those of other ferroelectric materials predicted from first-principles methods[11]. The derived spontaneous out-of-plane $P_{s\perp}$ is $-47.9$ μC/cm$^2$, which agrees well with STEM-HAADF results for unstrained BFO membranes ($-54.3 \pm 7.54$ μC/cm$^2$). Additionally, the coefficient of the effective piezoelectricity is obtainable from this model yielding $\tilde{e}_{zxx} = -12.4\pm0.4$ C/m$^2$ for BFO from the slope of the linear fitting between the polarization gradient and strain gradient (Fig. 2c).

With these coefficients of the effective piezoelectricity and flexoelectricity for BFO, we are able to calculate the contributions to the total polarization associated with both piezoelectricity and flexoelectricity measured in the external layer (the layer where maximum polarization exists) with different strain gradients (Fig. 2d). Both piezoelectricity and flexoelectricity contribute significantly to the total polarization in addition to the spontaneous polarization $P_{s\perp}$ in bent BFO. In particular, when the strain gradient reaches $3.7\times10^7$ m$^{-1}$, flexoelectricity offers more than 50% enhancement of the total polarization, equivalent to nearly three times that of spontaneous polarization. Therefore, because a bent membrane is capable of accommodating huge strain gradients, the induced flexoelectric polarization can be large enough to modulate the localized polarization.

With the absence of piezoelectricity, the bent STO has a uniform distribution in polarization across the membrane (Fig. 2a), unlike the bent BFO for which the distribution varies linearly with strain across the membrane arising from its piezoelectricity.

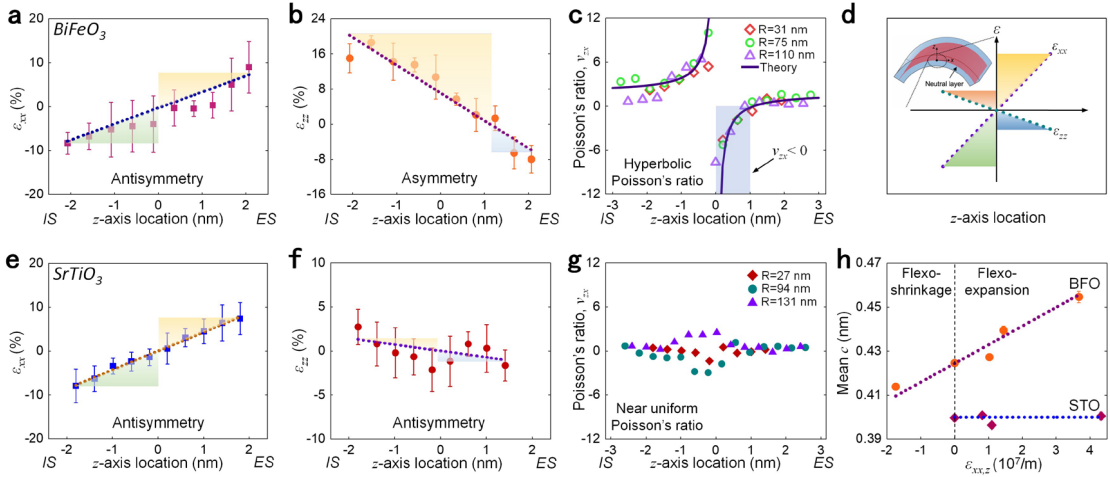

**FIG. 3. Mechanical behavior of bent BFO and STO membranes. a,** Antisymmetric in-plane and (**b**) asymmetric out-of-plane strain distributions in bent BFO (R=31 nm) along the thickness direction. **c,** Hyperbolic-like trend in Poisson's ratio bent BFO. **d,** Schematic of the theoretical antisymmetric in-plane and out-of-plane strain distributions in a bent membrane. **e-f,** Antisymmetric in-plane and out-of-plane strain distributions, respectively, for bent STO (R=27 nm) along the thickness direction. **g,** A near-uniform Poisson's ratio in bent STO. **h,** Mean lattice spacing $c$ as a function of the strain gradient in BFO and STO.

The huge strain gradient in bent freestanding perovskites not only induces a giant polarization, but also drives an unusual "bending-expansion" behavior, as well as hyperbolic-like distribution in Poisson's ratio. Classical elastic bending theory assumes that the in-plane and out-of-plane strains have linear antisymmetric distributions across the membrane (Fig. 3d). Indeed, the distribution of the in-plane strain $\varepsilon_{xx}$ in BFO and STO (Figs. 3a, e), as well as the out-of-plane strain $\varepsilon_{zz}$ in STO (Figs. 3f), agree well with this theory despite the thickness of these membranes being only several nanometers. The in-plane strain gradient $\varepsilon_{xx,z}$ also approximately obeys an inverse linear relationship with the radius of curvature. However, the out-of-plane strain distribution in BFO is found to be significantly asymmetric. (Figs. 3b). In consequence, this unusual distribution induces an abnormal change in membrane thickness under bending (herein, referred to as flexoexpansion or flexoshrinkage). The mean lattice spacing $c$ in bent BFO indeed increases (flexoexpansion) under a positive strain gradient and decreases (flexoshrinkage) under a negative one (Fig. 3h), but remains constant in bent STO. This change in thickness is proportional to the strain gradient and the overall thickness of BFO increases by 7.1% as the strain gradient reaches $3.7\times10^7$ m$^{-1}$. In addition to flexoexpansion or flexoshrinkage, an abnormal Poisson's ratio is found that follows a hyperbolic-like trend bent BFO (Fig. 3c) that exhibits a sudden change near the neutral layer. For example, bent BFO with R=31 nm has a Poisson's ratio of ~ $\pm10$ near the neutral layer, whereas its bulk value is only ~0.4[30]. To the best of our knowledge, this large change in Poisson's ratio and its hyperbolic-like trend has never been observed in single phase solids. However, surprisingly, this hyperbolic-like trend does not appear in bent STO (Fig. 3g).

To explain both the flexoexpansion/flexoshrinkage and the hyperbolic-like Poisson's ratio in freestanding BFO, we developed a linear electromechanical model. Expressions for the thickness of the bent membrane $h$ and Poisson's ratio $v_{zx}$ are as follows:

$$h = (A \cdot \varepsilon_{xx,z} + 1)h_0, \tag{1}$$

$$v_{zx} = -\frac{A}{z} - \frac{\varepsilon_{zz,z}}{\varepsilon_{xx,z}}, \tag{2}$$

$$A = \frac{d_{zxx}F_{zzzz} - d_{zzz}F_{zxxz}}{s_{xxxx}k_{zz} - d_{zxx}^2}, \tag{3}$$

where $h_0$ denotes the thickness of the flat membrane; $S_{ijkl}$, $d_{ijk}$, $F_{ijkl}$, and $k_{ij}$ denote the elastic compliance, piezoelectric, flexoelectric, and dielectric tensor, respectively. From equations (1) and (2), the thickness depends linearly on the strain gradient, and Poisson's ratio displays a hyperbolic trend. These abnormal trends exhibit a dependence on coefficient $A$, which is nonzero only when the material manifests piezoelectric and flexoelectric effects simultaneously. Our model indicates that the interplay between flexoelectricity and piezoelectricity provides a biased electromechanical out-of-plane strain. This explains why BFO shows a flexoexpansion effect and hyperbolic Poisson's ratio, but STO does not due to its lack of a piezoelectric effect.

A value for the linear coefficient $A$ of 2.0±0.4 nm for BFO was obtained by fitting equation (1) to the data (Fig. 3h). This value of $A$ = 2.0 nm indicates that the thickness varies by 2.0% per $1\times10^7\,\text{m}^{-1}$ of strain gradients and also explains why flexoexpansion has never been observed at the macroscopic scale, on which the strain gradient normally is only of order 10 m$^{-1}$. Using this $A$ value, the hyperbolic-like Poisson's ratio is well reproduced from this model (Fig. 3c).

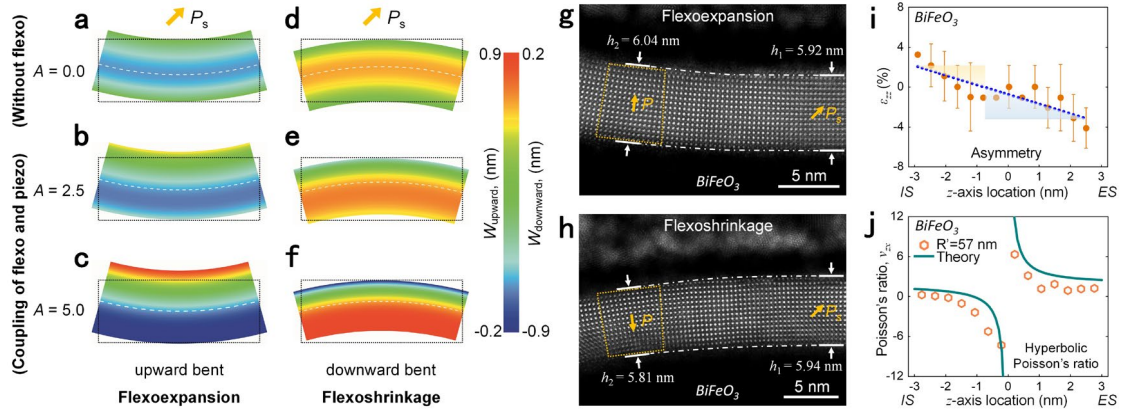

**FIG. 4. Flexoexpansion and flexoshrinkage effects in piezoelectric membranes. a-f,** Simulated bending deformation of piezoelectric membranes (5 nm thickness) with different bending expansion coefficients A=0.0 (**a**, **d**), 2.5 (**b**, **e**), and 5.0 (**c**, **f**) when bent upward (**a-c**) and downward (**d-f**), given that the out-of-plane spontaneous polarization (yellow arrow) points upwards. **g-h,** The flexoexpansion and flexoshrinkage effects observed in upward (**g**) and downward (**h**) bent BFO, respectively. **i-j,** Asymmetric out-of-plane strain distribution and hyperbolic-like trend in Poisson's ratio along the thickness direction measured in downward-bent BFO (R=57 nm) shown in (**h**).

When the membrane is bending in the $[00\bar{1}]$ direction (i.e., positive $z$ direction), the sign of the strain gradient $\varepsilon_{xx,z}$ is reversed as the strain decreases along the $[00\bar{1}]$ direction. Therefore, in accordance with equation (1), the thickness of the membrane narrows (flexoshrinkage). Both flexoexpansion and flexoshrinkage can be predicted from a theoretical perspective (Fig. 4a-f). In the experiment, we indeed, also found flexoshrinkage occurring in an oppositely bent BFO membrane (Figs. 4h), thereby validating our models. Interestingly, this asymmetrically distributed out-of-plane strain, as well as the hyperbolic-like Poisson's ratio (Fig. 4i, j), are inversely symmetric with those for upwardly bent membranes (Fig. 3b, c). The expansion or shrinkage across the membranes in different bending directions indicates that the polar freestanding oxides possess an asymmetric bending rigidity, which must be taken into account in future studies and applications.

The freestanding perovskite oxides exhibit an exceptional flexibility and capability to accommodate a giant strain gradient. The flexoelectric response becomes so predominant in comparison with piezoelectricity at atomic scale, that strain gradient engineering offers another path toward manipulating electrical and mechanical behaviors in these strongly correlated two-dimensional materials as future potential building blocks in multifunctional flexible electronics[31-33] and nanomachines[2]. We also discovered in experiments that the interplay between flexoelectricity and piezoelectricity leads to a hyperbolic-like trend in Poisson's ratio and an asymmetric change in bending-thickness with respect to the sign of the spontaneous polarization in low-dimensional polar material. This flexoexpansion enhances the flexural stiffness, which is related to the membrane thickness, and this behavior is reversible by switching the polarization that induces flexoshrinkage. The thickness-dependent hyperbolic-like Poisson's ratio may also induce the similar behavior as mechanical-gradient materials, even though their composition is homogenous. These unusual mechanical properties are expected to make BFO and other ferroelectric membranes effective smart mechanical materials[18] and strongly influence nano-mechanical performances regarding, for example, vibration, fracture, and wrinkling modes that play a crucial role in applications of nano-electromechanical systems and three-dimensional self-assembled nano-structures[34,35].

# References

1.  Kalaee, M. et al. Quantum electromechanics of a hypersonic crystal. *Nat Nanotechnol* **14**, 334-339 (2019).

2.  Ghatge, M., Walters, G., Nishida, T. & Tabrizian, R. An ultrathin integrated nanoelectromechanical transducer based on hafnium zirconium oxide. *Nature Electronics* **2**, 506-512 (2019).

3.  Choi, K. J. et al. Enhancement of Ferroelectricity in Strained BaTiO$_3$ Thin Films. *Science* **306**, 1005 (2004).

4.  Tang, Y. L. et al. Observation of a periodic array of flux-closure quadrants in strained ferroelectric PbTiO$_3$ films. *Science* **348**, 547 (2015).

5.  Das, S. et al. Observation of room-temperature polar skyrmions. *Nature* **568**, 368-372 (2019).

6.  Haeni, J. H. et al. Room-temperature ferroelectricity in strained SrTiO$_3$. *Nature* **430**, 758-761 (2004).

7.  Zhang, L. et al. Giant polarization in super-tetragonal thin films through interphase strain. *Science* **361**, 494 (2018).

8.  Wang, B., Gu, Y., Zhang, S. & Chen, L.-Q. Flexoelectricity in solids: Progress, challenges, and perspectives. *Progress in Materials Science* **106** (2019).

9.  Tang, Y. L. et al. Giant linear strain gradient with extremely low elastic energy in a perovskite nanostructure array. *Nat Commun* **8**, 15994 (2017).

10. Lee, D. et al. Giant Flexoelectric Effect in Ferroelectric Epitaxial Thin Films. *Physical Review Letters* **107** (2011).

11. Zubko, P., Catalan, G. & Tagantsev, A. K. Flexoelectric Effect in Solids. *Annual Review of*

*Materials Research* **43**, 387-421 (2013).

12. Lu, D. et al. Synthesis of freestanding single-crystal perovskite films and heterostructures by etching of sacrificial water-soluble layers. *Nature materials* **15**, 1255-1260 (2016).

13. Ji, D. et al. Freestanding crystalline oxide perovskites down to the monolayer limit. *Nature* **570**, 87-90 (2019).

14. Dong, G. et al. Super-elastic ferroelectric single-crystal membrane with continuous electric dipole rotation. *Science* **366**, 475 (2019).

15. Eerenstein, W., Mathur, N. D. & Scott, J. F. Multiferroic and magnetoelectric materials. *Nature* **442**, 759-765 (2006).

16. Zhang, F. et al. Modulating the Electrical Transport in the Two-Dimensional Electron Gas at $LaAlO_3/SrTiO_3$ Heterostructures by Interfacial Flexoelectricity. *Phys Rev Lett* **122**, 257601 (2019).

17. Hwang, J. et al. Tuning perovskite oxides by strain: Electronic structure, properties, and functions in (electro)catalysis and ferroelectricity. *Materials Today* **31**, 100-118 (2019).

18. Cordero-Edwards, K. et al. Ferroelectrics as Smart Mechanical Materials. *Adv Mater* **29** (2017).

19. Bhaskar, U. K. et al. Flexoelectric MEMS: towards an electromechanical strain diode. *Nanoscale* **8**, 1293-1298 (2016).

20. Cordero-Edwards, K. et al. Flexoelectric Fracture-Ratchet Effect in Ferroelectrics. *Phys Rev Lett* **122**, 135502 (2019).

21. Tang, Y. L. et al. Atomic-scale mapping of dipole frustration at 90 degrees charged domain walls in ferroelectric $PbTiO_3$ films. *Sci Rep* **4**, 4115 (2014).

22. Lu, H. et al. Mechanical Writing of Ferroelectric Polarization. *Science* **336**, 59 (2012).

23. Yang, M.-M., Kim, D. J. & Alexe, M. Flexo-photovoltaic effect. *Science* **360**, 904 (2018).

24. Abrahams, S. C., Kurtz, S. K. & Jamieson, P. B. Atomic Displacement Relationship to Curie Temperature and Spontaneous Polarization in Displacive Ferroelectrics. *Physical Review* **172**, 551-553 (1968).

25. Gao, P. et al. Atomic-Scale Measurement of Flexoelectric Polarization at $SrTiO_3$ Dislocations. *Phys Rev Lett* **120**, 267601 (2018).

26. Xie, L. et al. Giant Ferroelectric Polarization in Ultrathin Ferroelectrics via Boundary-Condition Engineering. *Adv Mater* **29** (2017).

27. Sun, Y. et al. Subunit cell–level measurement of polarization in an individual polar vortex. *Science Advances* **5**, eaav4355 (2019).

28. Zhang, J. X. et al. Microscopic origin of the giant ferroelectric polarization in tetragonal-like $BiFeO_3$. *Phys Rev Lett* **107**, 147602 (2011).

29. Wang, H. et al. Direct Observation of Huge Flexoelectric Polarization around Crack Tips. *Nano Letters* **20**, 88-94 (2019).

30. Dong, H. et al. Elastic properties of tetragonal $BiFeO_3$ from first-principles calculations. *Applied Physics Letters* **102**, 182905 (2013).

31. Qi, Y. et al. Enhanced piezoelectricity and stretchability in energy harvesting devices fabricated from buckled PZT ribbons. *Nano letters* **11**, 1331-1336 (2011).

32. Deng, Q., Kammoun, M., Erturk, A. & Sharma, P. Nanoscale flexoelectric energy harvesting. *International Journal of Solids and Structures* **51**, 3218-3225 (2014).

33. Zhang, S., Liu, K., Xu, M. & Shen, S. A curved resonant flexoelectric actuator. *Applied Physics Letters* **111**, 082904 (2017).

34. Cheng, X. & Zhang, Y. Micro/Nanoscale 3D Assembly by Rolling, Folding, Curving, and Buckling Approaches. *Adv Mater* **31**, e1901895 (2019).

35. Chen, Z. et al. Mechanical Self-Assembly of a Strain-Engineered Flexible Layer: Wrinkling, Rolling, and Twisting. *Physical Review Applied* **5** (2016).

## Acknowledgments

This work was supported by the National Basic Research Program of China (grant 2015CB654901), the National Natural Science Foundation of China (grant: 11874199, 11774153, 1861161004), the International Cooperation and Exchange Program by NSFC (11911530174) and the Fundamental Research Funds for the Central Universities (020514380224, 14380167). J.W.H acknowledges support from the National Science Foundation of China (grant 11572040), Beijing Natural Science Foundation (Z190011) and the Technological Innovation Project of Beijing Institute of Technology. C.A. acknowledges support from the U.S. Department of Energy, Office of Basic Energy Science, Division of Materials Science and Engineering (DE-SC0014430). D.X.J. is supported by Program A for Outstanding Ph.D. candidate of Nanjing University (grant 201901A014). Y.Z.L. is supported by Graduate Technological Innovation Project of Beijing Institute of Technology (grant 2019CX20002).

## Author contributions

P.W. and X.Q.P supervised the STEM characterizations. Y.F.N. and Z.B.G. supervised the synthesis of epitaxial and freestanding films. J.W.H and X.Y.W supervised theoretical analysis and phase-field simulations. S.H.C. carried out SEM observation and prepared the TEM cross-sectional samples via FIB. S.H.C., D.X.J., C.C.Z. and M.G. carried out STEM experiments. S.H.C., Y.Z.L., D.X.J., Y.F.W. and S.G. carried out data analysis. Y.Z.L. and J.W.H. carried out theoretical analysis. C.Q.G. carried out phase-field simulations. D.X.J., Y.P.Z. and L.H. grew and transferred the freestanding perovskite oxides. S.H.C., Y.Z.L., P.W., J.W.H., X.Q.P., Y.F.N., D.X.J., C.A. and L.H. wrote and edited the manuscript. All authors discussed the data and contributed to the manuscript.